\documentclass[%
 reprint,
 amsmath,amssymb,
 aps,
pra,
floatfix,
]{revtex4-2}
\usepackage{upgreek}
\usepackage{graphicx}
\usepackage{dcolumn}
\usepackage{bm}
\usepackage{marvosym}
\usepackage{float}

\begin{document}


\title{Tightly-confined and long Z-cut lithium niobate waveguide with ultralow-loss}

\author{Yan Gao}
\author{Yi Sun}
\author{Israel Rebolledo-Salgado}
\author{Raphaël Van Laer}
\author{Victor Torres-Company}
\author{Jochen Schröder\textsuperscript{\Letter}}
\email{jochen.schroeder@chalmers.se}

 \affiliation{\textsuperscript{1}Department of Microtechnology and Nanoscience (MC2), Chalmers University of Technology, Göteborg, Sweden\\}





\begin{abstract}
Lithium niobate (LN) is a promising material for future complex photonic-electronic circuits, with wide applications in fields like data communications, sensing, optical computation, and quantum optics. There was a great step toward LN photonic integrated circuits (PICs) with the development of dry etching for low-loss LN on insulator (LNOI) waveguides. However, the versatility of the LN waveguide platform for applications like $\chi^3$ nonlinear devices and passive phase sensitive components, has not been fully utilized. The main challenges are the difficulty of making highly confined ultralow-loss waveguides and overcoming the strong material birefringence. Here, we developed a fabrication technology for an ultralow-loss, tightly-confined, dispersion-engineered LN waveguide. We demonstrated an ultra-low propagation loss of 5.8 dB/m in a decimeter-long LN spiral waveguide. We focused on Z-cut LN waveguides with TE mode to avoid the material birefringence. Aiming for $\chi^3$ nonlinear applications, we demonstrated the first all normal-dispersion (ANDi) based coherent octave-spanning supercontinuum frequency comb in integrated LN waveguide. Our ultralow-loss Z-cut LN long waveguide might be useful in on-chip narrow linewidth lasers, optical delay lines, and parametric amplifiers.
\end{abstract}

\maketitle


\noindent
\textbf{1. Introduction
}

 Photonic integrated circuits \cite{thomson2016roadmap} (PICs) enable on-chip light generation, manipulation, and detection. Via integration and miniturization, PICs shows great potential for realizing low-cost and scalable optical systems in fields like data communication, bio-chemical sensing, and optical computation. In recent decades, different material platforms have been investigated, including silicon (Si) \cite{jalali2006silicon}, indium phosphide (InP) \cite{smit2019past}, silicon nitride (SiN$_x$) \cite{xuan2016high, ji2017ultra, liu2018ultralow, ye2019high}, aluminum gallium arsenide (AlGaAs) \cite{pu2016efficient, chang2020ultra}, aluminium nitride (AlN) \cite{jung2013electrical}, silicon carbide (SiC) \cite{guidry2020optical}, and lithium niobate (LN) \cite{zhang2017monolithic, he2019self, gong2019soliton, gao2023compact}.
 
 Recently, LN has attracted large research interest due to its unique properties that can simultaneously provide electro-optic (EO), nonlinear, acousto-optic effects. Moreover, it exhibits a broad optical transparency window, a high refractive-index-contrast, and ultra-low material losses \cite{zhu2021integrated,qi2020integrated,honardoost2020rejuvenating,weis1985lithium}. Photonic applications in LN
such as low drive voltage high-speed EO modulators \cite{wang2018integrated,he2019high}, Kerr and EO frequency combs \cite{he2019self,zhang2019broadband}, ultra-efficient frequency converters \cite{lu2019periodically,chen2019ultra}, squeezed light sources \cite{nehra2022few}, photon pair sources \cite{zhao2020high} and parametric optical oscillators \cite{lu2021ultralow, mckenna2022ultra} have been demonstrated.

Long waveguides with ultralow losses are crucial for a host of chip-scale applications, such as narrow linewidth lasers \cite{xiang2020narrow}, optical delay lines \cite{lee2012ultra, hong2021ultralow, ji2019chip}, and nonlinear optical processes like parametric amplifiers \cite{ye2021overcoming, kazama2021over}. Such waveguides have been built in dielectric materials like Si$_3$N$_4$ \cite{bauters2011ultra, ye2021overcoming, liu2021high} and SiO$_2$ \cite{lee2012ultra} with losses in the dB/m regime. For LNOI, however, it is a bigger challenge and has so far remained elusive because of the waveguide fabrication difficulty and strong material birefringence.

In this work, we leverage recent advances in fully etched LN waveguides \cite{gao2023compact} and further develop the methodology to achieve ultra-low loss strongly confined Z-cut spiral-waveguides. We demonstrated propagation losses down to 5.8 dB/m in a decimeter-long spiral waveguide and give a detailed discussion about the fabrication steps. The ultra-low loss combined with a Z-cut configuration allows us to reach decimeter long interaction lengths, enabling octave-spanning supercontinuum in an all-normal-dispersion LN waveguide for the first time, to the best of our knowledge.

\noindent
\textbf{2. Z-cut LN long waveguide platform}

\begin{figure}[!b]
\centering\includegraphics[width=0.48\textwidth]{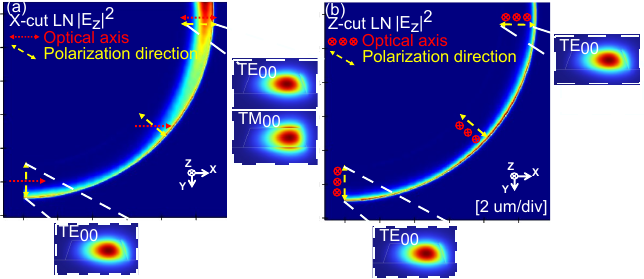}
\caption{(a) The simulated optical field (Z component) distribution in a X-cut LN bent waveguide. The input is TE$_{00}$ mode. After propagation, partial energy transferred to TM$_{00}$ mode due to the index anisotropy. (b) The same field profile but in a Z-cut LN bent waveguide, where there is no energy transfer from TE$_{00}$ to TM$_{00}$.}
\label{fig1_all}
\end{figure}

\begin{figure*}
\centering\includegraphics[width=0.8\textwidth]{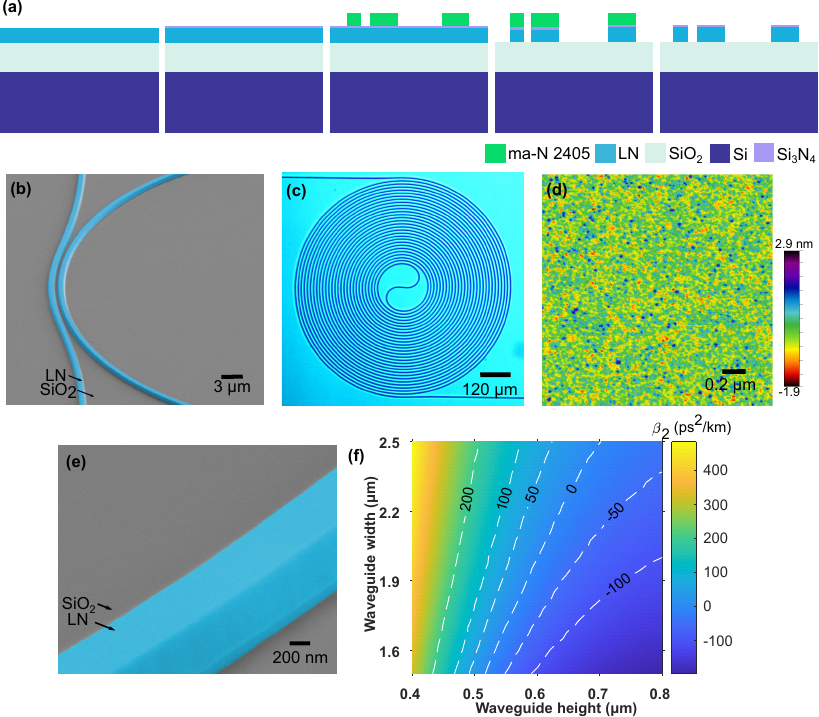}
\caption{(a) Fabrication flow chart for fully etched LN waveguide. (b) SEM image of LN ring resonators and (e) zoomed in SEM image for waveguide sidewall, the figure was post-colored to highlight LN waveguide. (c) Top view microscope image for a spiral waveguide in one writing field. (d) AFM measurement for the surface roughness of Si$_3$N$_4$ sputtered on LN, the measured surface RMS roughness is 0.5 nm. (f) Simulated dispersion (FEM, COMSOL multiphysics) of the fully etched LN waveguide with the bending radius of 200 $\upmu$m.}
\label{fig2_all}
\end{figure*}

LN is a well-known anisotropic material, in which the nonlinear and EO coefficients vary largely in different directions. To take advantage of the largest EO (r$_3$$_3$) or second order nonlinear coefficient (d$_3$$_3$), most work has focused on the TE polarization in X-cut LN waveguides. However, the strong birefringence limits the waveguide layout to a single direction. Such configuration will also make the in-plane optical index to experience large changes ($\Delta$n = 0.08) with waveguide direction because of the birefringence. The resulting direction-dependent index will cause serious intermode-crosstalk \cite{pan2020compact} even in a single-moded bent waveguide. The index anisotropy will increase the design complexity and special consideration is required for phase sensitive or dispersion components such as arrayed waveguide grating (AWG) \cite{yi2024anisotropy}. As showed in Fig. 1(a), for an X-cut bent waveguide, the light mode will be distorted by the index anisotropy, and energy of the input TE$_0$$_0$ mode will transfer to the TM$_0$$_0$ mode. However, the mode will maintain the same energy when propagating in a Z-cut bent waveguide. 

\begin{figure*}[htb]
\centering\includegraphics[width=0.65\textwidth]{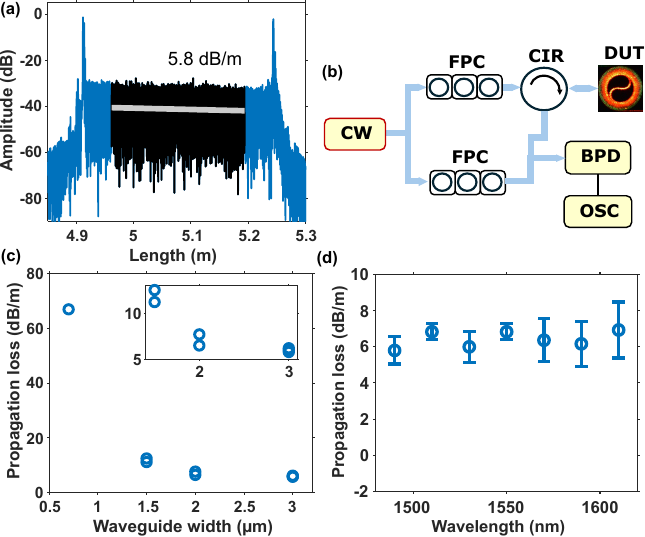}
\caption{(a) OFDR measurement results for a 15 cm long waveguide. (b) Experimental setup. (c) Measured losses with different waveguide width. (d) Measured losses with different wavelength.}
\label{fig3}
\end{figure*}

Below we give a detailed explanation of our fabrication process to obtain ultra-low loss tightly-confined LN waveguides.

The samples for the waveguide fabrication are diced from a 4 inch commercial LN wafer (NANOLN). The wafer includes 600 nm thin film LN on top, 525 $\upmu$m thick silicon substrate at the bottom, and 4.7 $\upmu$m thermally oxidized SiO$_2$ in between. Fig. 2(a) shows the fabrication flow. The sample is first prepared via solvent cleaning (acetone, IPA) and then via standard cleaning (SC1, 29\% NH$_3$ : 30\% H$_2$O$_2$ : H$_2$O = 1:1:5). To minimize the effect from the SC1 solution on the LN, we limit the SC1 time to 2 minutes only. The sample is then sent to the vacuum sputtering tool (FHR, MS150), and around 5 nm Si$_3$N$_4$ is deposited on the top of the LN, to promote the adhesion between LN and resist. Using atomic force microscopy (AFM), we measure the surface roughness of the sample after Si$_3$N$_4$ deposition, and the measured root mean square roughness (RMS) is 0.5 nm. The surface roughness is slightly increased compared to the pure LN surface (around 0.3 nm). However, the increased top surface roughness does not contribute significantly to the waveguide losses according to our measurement results, and we believe that the losses are still mainly governed by the waveguide sidewall roughness.

Negative tone electro-beam lithography (EBL) resist ma-N 2405 is adopted for its high resolution, dry etch resistance, and more importantly, good thermal stability. The pattern is first defined on the resist via 100-kV EBL system (Raith, EBPG 5200), in which multipass exposure is used to reduce sidewall roughness. A beam step size (BSS) of 3 nm was chosen and the delivered doses were 230 $\upmu$C/cm$^2$. The dose will influence the sidewall roughness and should be optimized for different fabrication flows. Moreover, resistance to ion milling is important because of the pure physical etching process and the deep etching depth here. We found that a larger dose tends to give better resistance during dry etching. As a result, the dose should be optimized by considering both the roughness and the resist resistance.

After exposure and development, the sample is then dry-etched via reactive ion beam etching (IBE, Oxford Ionfab 300 Plus) with only Ar$^+$ plasma. The Ar$^+$ gas flow used for the plasma generation is 6 sccm. A higher concentration of Ar$^+$ plasma will lead to a higher etching rate. The generated Ar$^+$ plasma will be accelerated via electric field and directly bombards the LN surface. The pure physical etching process results in poor selectivity (around 1.4). The measured etching rate of LN is around 14 nm/min and 45 min is required to achieve slightly overetching for 600 nm LN. During the etching process, especially for our fully etched waveguides, we observe strong thermal accumulation that can cause local burning of the resist even with helium cooling. We therefore adopt a strategy to keep the temperature at a certain low level: 1) Choose a thermally stable resist (ma-N 2405); 2) Use thermal release tape between sample and carrier wafer; 3) Perform multiple etching steps to allow cooling of the sample. The sidewall angle of the waveguides is estimated to be around 70 degrees which is due to the lateral etching of resist during the physical etching. Finally, the sample is sent for another run of solvent and standard cleaning to remove the remaining resist and by-products.

\begin{figure*}
\centering\includegraphics[width=0.7\textwidth]{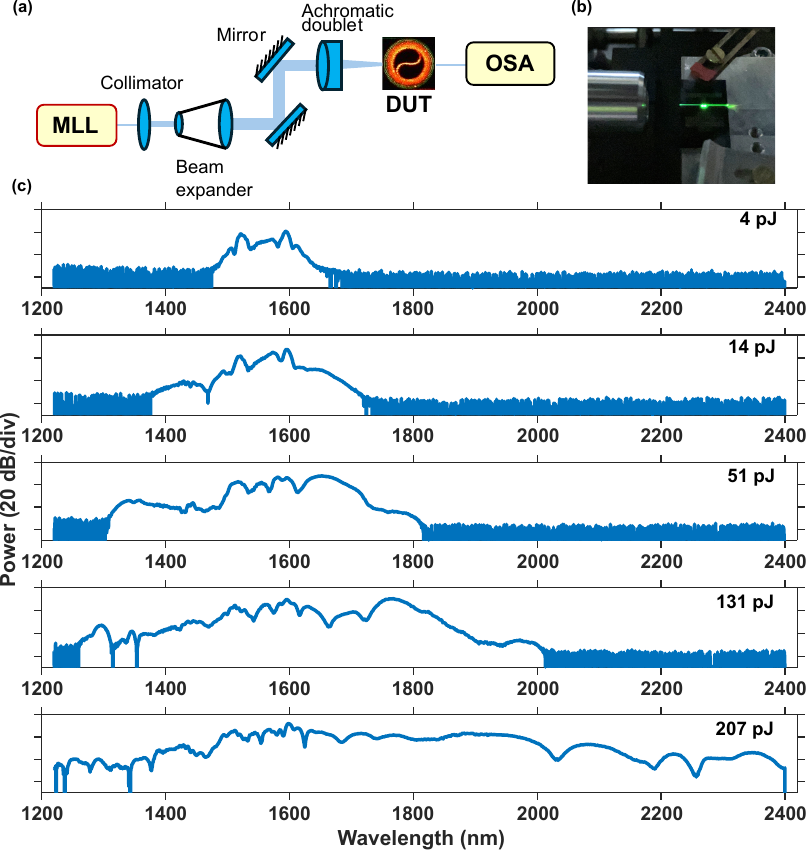}
\caption{(a) Experimental setup for supercontinuum generation, the light is free-space couled onto the chips. (b) Picture for the measured setup. (c) The measured spectrum when pump with different power.}
\label{fig4}
\end{figure*}

\noindent
\textbf{3. Losses characterization}


     

We fabricated long spiral waveguides to characterize the propagation losses. Fig. 2(c) shows a microscope image for a fabricated spiral waveguide.

The spiral shape reduces the footprint for the entire device. We minimize stiching errors (and thus extra losses), by intentionally fitting each spiral unit into a single writing file (1 $\times$ 1 mm$^2$) \cite{ye2021overcoming}. Each spiral unit includes two centrosymmetric Archimedean spirals and an S-bend to connect them. Since any sharp change in curvature of a waveguide will introduce radiation loss and intermode coupling between different transverse modes, a smoothly varying S-bend is needed to connect the two Archimedean spirals. We consider a curves family in which the curvature is given in terms of a cubic polynomial of arc length '$s$' \cite{chen2012general, ye2022integrated}:
\begin{equation}\label{...}
\kappa(s) = a_0 + a_1s + a_2s^2+a_3s^3
\end{equation}
The coefficients $a_0$, $a_1$, $a_2$, $a_3$ can be solved by using the boundary conditions at the initial and final connection points with physical position, the tangent, the curvature and the change of the curvature.

Optical frequency-domain reflectometry (OFDR) based on Rayleigh scattering \cite{liang2021comprehensive} is used to characterize the propagation losses of our waveguide. The measurement setup is shown in Fig. 3(b). It includes a sweeping CW laser, fiber delay path, and a photo-detector. By applying an inverse-Fourier transform of the temporal distribution of the reflected power from the spiral waveguide, the reflected power as a function of time delay/length can be determined and the propagation loss can be obtained by linear fitting. Fig 3(a) shows OFDR measurement result for a 15 cm long spiral waveguide with the lowest loss, and the measured loss is 5.8 dB/m. To the best of our knowledge, this constitutes the lowest loss measured directly in a decimeter-long LN waveguide. The measured loss corresponds to a mean value over the wavelength range of 1480-1620 nm. To further characterize the loss at different wavelengths, we considered 3 different devices with the same waveguide width of 3 $\upmu$m and processed the data to obtain the loss at different wavelengths. The measured results are consistent with each other, and all devices exhibit a similar loss over the full measured wavelength range. Propagation losses with different waveguide widths from 700 nm to 3 $\upmu$m were also measured and are shown in Fig. 3(c). The propagation loss is around 0.67 dB/cm in the single mode regime (700 nm) and decreases dramatically for multi-mode waveguides, as the mode will be more confined and located in the central region and experiences less interaction with the sidewall roughness. Ultra-low propagation losses down to a few dB/m has been demonstrated here for reasonable widths multi-mode waveguides.

\noindent
\textbf{4. Nonlinear applications based on $\chi ^3$}

We study supercontinuum generation in an all-normal dispersion LN waveguide with geometry of 2.7 $\times$ 0.6 $\upmu$m$^2$ and length of 30 cm. The device used for the experiment includes 6 cascaded spiral units with around 5 cm length for each. The measurement setup is shown in Fig 4. (a). A 50 femtosecond mode lock laser (MLL) with a center wavelength of 1560 nm and repetition rate of 250 MHz is used to pump the device. The laser beam is first collimated and expanded to obtain a diffraction limited spot size. The mirrors are used to align the height of the beam and the achromatic doublet is used to focus the beam onto the waveguide facet. A half-wave plate is used to tune the beam polarization to couple the light into the waveguide's TE mode. The estimated input coupling loss is around 10 dB. We attribute the large coupling loss to the suboptimal overlap between waveguide and the focused beam spot. The light is coupled from the waveguide to a lensed fiber with beam diameter of 2.5 $\upmu$m. After the lensed fiber, we use a single mode fiber connected to an OSA (wavelength range of 1200 nm - 2400 nm) to measure the spectrum.

We carried out measurements with varying optical input power using a series of free space optical density filters. As shown in Fig. 4(c), the spectrum broadens when the on-chip pump power is increased. The spectrum reaches one octave spanning at a pump pulse energy of around 207 pJ. Unlike anomalous dispersion supercontinuum, the coherence of the broadened pulse in the ANDi waveguides is maintained over relatively long propagation lengths and high energy pulses \cite{finot2008beneficial,dudley2002coherence,rebolledo2022coherent}. The maximum spectrum bandwidth is limited by the available MLL power and the suboptimal coupling losses from free space to chip.


\noindent
\textbf{5. Conclusion}

In conclusion, a LN waveguide fabrication technology has been developed to realize a fully-etched strip LN waveguide with advantages for simultaneously achieving ultra-low propagation losses, strong light confinement and dispersion engineering. We focused on Z-cut LN to avoid the material anisotropy, but the fabrication technology is not limited to only Z-cut LN. Utilizing the developed LN waveguide, we showed an ultralow-loss down to a few dB/m by using a 15-cm long spiral waveguide. Utilizing the long waveguide, we demonstrate an octave-spanning supercontinuum in the all normal-dispersion regime. Our ultralow loss long LN waveguide is desirable for chip-scale narrow linewidth lasers \cite{xiang2020narrow}, low loss optical delay lines \cite{lee2012ultra, hong2021ultralow, ji2019chip}, and parametric amplification \cite{ye2021overcoming, kazama2021over}. Besides, the waveguide platform could also be used for situations where densely integrated optical components are required, such as large-scale optical switching arrays \cite{seok2019wafer}, integrated photonic LiDAR \cite{zhang2022large}, and optical artificial intelligence \cite{zhang2021optical}.

\vspace{0.4cm}
\noindent
$\mathbf{Funding:}$ Swedish Research council (VR-2017-05157, VR-2021-04241);

\vspace{0.2cm}
\noindent
 $\mathbf{Acknowledgments:}$ The authors thank cleanroom staff from Myfab at Chalmers Nanofabrication Laboratory for discussion and training.

\vspace{0.2cm}
\noindent
 $\mathbf{Disclosures:}$ The authors declare no conflicts of interest.

\vspace{0.2cm}
\noindent
$\mathbf{Data}$ $\mathbf{availability:}$ Data underlying the results presented in this paper are available in Ref. \cite{gao_2025_14606023}.

\bibliography{sample}

\begin{thebibliography}{48}%
\makeatletter
\providecommand \@ifxundefined [1]{%
 \@ifx{#1\undefined}
}%
\providecommand \@ifnum [1]{%
 \ifnum #1\expandafter \@firstoftwo
 \else \expandafter \@secondoftwo
 \fi
}%
\providecommand \@ifx [1]{%
 \ifx #1\expandafter \@firstoftwo
 \else \expandafter \@secondoftwo
 \fi
}%
\providecommand \natexlab [1]{#1}%
\providecommand \enquote  [1]{``#1''}%
\providecommand \bibnamefont  [1]{#1}%
\providecommand \bibfnamefont [1]{#1}%
\providecommand \citenamefont [1]{#1}%
\providecommand \href@noop [0]{\@secondoftwo}%
\providecommand \href [0]{\begingroup \@sanitize@url \@href}%
\providecommand \@href[1]{\@@startlink{#1}\@@href}%
\providecommand \@@href[1]{\endgroup#1\@@endlink}%
\providecommand \@sanitize@url [0]{\catcode `\\12\catcode `\$12\catcode `\&12\catcode `\#12\catcode `\^12\catcode `\_12\catcode `\%12\relax}%
\providecommand \@@startlink[1]{}%
\providecommand \@@endlink[0]{}%
\providecommand \url  [0]{\begingroup\@sanitize@url \@url }%
\providecommand \@url [1]{\endgroup\@href {#1}{\urlprefix }}%
\providecommand \urlprefix  [0]{URL }%
\providecommand \Eprint [0]{\href }%
\providecommand \doibase [0]{https://doi.org/}%
\providecommand \selectlanguage [0]{\@gobble}%
\providecommand \bibinfo  [0]{\@secondoftwo}%
\providecommand \bibfield  [0]{\@secondoftwo}%
\providecommand \translation [1]{[#1]}%
\providecommand \BibitemOpen [0]{}%
\providecommand \bibitemStop [0]{}%
\providecommand \bibitemNoStop [0]{.\EOS\space}%
\providecommand \EOS [0]{\spacefactor3000\relax}%
\providecommand \BibitemShut  [1]{\csname bibitem#1\endcsname}%
\let\auto@bib@innerbib\@empty
\bibitem [{\citenamefont {David}\ \emph {et~al.}(2016)\citenamefont {David}, \citenamefont {Aaron}, \citenamefont {John~E}, \citenamefont {Komljenovic}, \citenamefont {Graham~T}, \citenamefont {Laurent}, \citenamefont {Delphine}, \citenamefont {Eric}, \citenamefont {Léopold}, \citenamefont {Jean-Marc}, \citenamefont {Jean-Michel}, \citenamefont {Jens~H}, \citenamefont {Dan-Xia}, \citenamefont {Frédéric}, \citenamefont {Peter}, \citenamefont {Goran~Z},\ and\ \citenamefont {M}}]{thomson2016roadmap}%
  \BibitemOpen
  \bibfield  {author} {\bibinfo {author} {\bibfnamefont {T.}~\bibnamefont {David}}, \bibinfo {author} {\bibfnamefont {Z.}~\bibnamefont {Aaron}}, \bibinfo {author} {\bibfnamefont {B.}~\bibnamefont {John~E}}, \bibinfo {author} {\bibfnamefont {T.}~\bibnamefont {Komljenovic}}, \bibinfo {author} {\bibfnamefont {R.}~\bibnamefont {Graham~T}}, \bibinfo {author} {\bibfnamefont {V.}~\bibnamefont {Laurent}}, \bibinfo {author} {\bibfnamefont {M.-M.}\ \bibnamefont {Delphine}}, \bibinfo {author} {\bibfnamefont {C.}~\bibnamefont {Eric}}, \bibinfo {author} {\bibfnamefont {V.}~\bibnamefont {Léopold}}, \bibinfo {author} {\bibfnamefont {F.}~\bibnamefont {Jean-Marc}}, \bibinfo {author} {\bibfnamefont {H.}~\bibnamefont {Jean-Michel}}, \bibinfo {author} {\bibfnamefont {S.}~\bibnamefont {Jens~H}}, \bibinfo {author} {\bibfnamefont {X.}~\bibnamefont {Dan-Xia}}, \bibinfo {author} {\bibfnamefont {B.}~\bibnamefont {Frédéric}}, \bibinfo {author} {\bibfnamefont {O.}~\bibnamefont {Peter}}, \bibinfo {author} {\bibfnamefont
  {M.}~\bibnamefont {Goran~Z}},\ and\ \bibinfo {author} {\bibfnamefont {N.}~\bibnamefont {M}},\ }\bibfield  {title} {\bibinfo {title} {Roadmap on silicon photonics},\ }\href@noop {} {\bibfield  {journal} {\bibinfo  {journal} {J. Opt.}\ }\textbf {\bibinfo {volume} {18}},\ \bibinfo {pages} {073003} (\bibinfo {year} {2016})}\BibitemShut {NoStop}%
\bibitem [{\citenamefont {Jalali}\ and\ \citenamefont {Fathpour}(2006)}]{jalali2006silicon}%
  \BibitemOpen
  \bibfield  {author} {\bibinfo {author} {\bibfnamefont {B.}~\bibnamefont {Jalali}}\ and\ \bibinfo {author} {\bibfnamefont {S.}~\bibnamefont {Fathpour}},\ }\bibfield  {title} {\bibinfo {title} {Silicon photonics},\ }\href@noop {} {\bibfield  {journal} {\bibinfo  {journal} {J. Light. Technol.}\ }\textbf {\bibinfo {volume} {24}},\ \bibinfo {pages} {4600} (\bibinfo {year} {2006})}\BibitemShut {NoStop}%
\bibitem [{\citenamefont {Smit}\ \emph {et~al.}(2019)\citenamefont {Smit}, \citenamefont {Williams},\ and\ \citenamefont {Van Der~Tol}}]{smit2019past}%
  \BibitemOpen
  \bibfield  {author} {\bibinfo {author} {\bibfnamefont {M.}~\bibnamefont {Smit}}, \bibinfo {author} {\bibfnamefont {K.}~\bibnamefont {Williams}},\ and\ \bibinfo {author} {\bibfnamefont {J.}~\bibnamefont {Van Der~Tol}},\ }\bibfield  {title} {\bibinfo {title} {Past, present, and future of inp-based photonic integration},\ }\href@noop {} {\bibfield  {journal} {\bibinfo  {journal} {Apl Photonics}\ }\textbf {\bibinfo {volume} {4}} (\bibinfo {year} {2019})}\BibitemShut {NoStop}%
\bibitem [{\citenamefont {Xuan}\ \emph {et~al.}(2016)\citenamefont {Xuan}, \citenamefont {Liu}, \citenamefont {Varghese}, \citenamefont {Metcalf}, \citenamefont {Xue}, \citenamefont {Wang}, \citenamefont {Han}, \citenamefont {Jaramillo-Villegas}, \citenamefont {Al~Noman}, \citenamefont {Wang} \emph {et~al.}}]{xuan2016high}%
  \BibitemOpen
  \bibfield  {author} {\bibinfo {author} {\bibfnamefont {Y.}~\bibnamefont {Xuan}}, \bibinfo {author} {\bibfnamefont {Y.}~\bibnamefont {Liu}}, \bibinfo {author} {\bibfnamefont {L.~T.}\ \bibnamefont {Varghese}}, \bibinfo {author} {\bibfnamefont {A.~J.}\ \bibnamefont {Metcalf}}, \bibinfo {author} {\bibfnamefont {X.}~\bibnamefont {Xue}}, \bibinfo {author} {\bibfnamefont {P.-H.}\ \bibnamefont {Wang}}, \bibinfo {author} {\bibfnamefont {K.}~\bibnamefont {Han}}, \bibinfo {author} {\bibfnamefont {J.~A.}\ \bibnamefont {Jaramillo-Villegas}}, \bibinfo {author} {\bibfnamefont {A.}~\bibnamefont {Al~Noman}}, \bibinfo {author} {\bibfnamefont {C.}~\bibnamefont {Wang}}, \emph {et~al.},\ }\bibfield  {title} {\bibinfo {title} {High-q silicon nitride microresonators exhibiting low-power frequency comb initiation},\ }\href@noop {} {\bibfield  {journal} {\bibinfo  {journal} {Optica}\ }\textbf {\bibinfo {volume} {3}},\ \bibinfo {pages} {1171} (\bibinfo {year} {2016})}\BibitemShut {NoStop}%
\bibitem [{\citenamefont {Ji}\ \emph {et~al.}(2017)\citenamefont {Ji}, \citenamefont {Barbosa}, \citenamefont {Roberts}, \citenamefont {Dutt}, \citenamefont {Cardenas}, \citenamefont {Okawachi}, \citenamefont {Bryant}, \citenamefont {Gaeta},\ and\ \citenamefont {Lipson}}]{ji2017ultra}%
  \BibitemOpen
  \bibfield  {author} {\bibinfo {author} {\bibfnamefont {X.}~\bibnamefont {Ji}}, \bibinfo {author} {\bibfnamefont {F.~A.}\ \bibnamefont {Barbosa}}, \bibinfo {author} {\bibfnamefont {S.~P.}\ \bibnamefont {Roberts}}, \bibinfo {author} {\bibfnamefont {A.}~\bibnamefont {Dutt}}, \bibinfo {author} {\bibfnamefont {J.}~\bibnamefont {Cardenas}}, \bibinfo {author} {\bibfnamefont {Y.}~\bibnamefont {Okawachi}}, \bibinfo {author} {\bibfnamefont {A.}~\bibnamefont {Bryant}}, \bibinfo {author} {\bibfnamefont {A.~L.}\ \bibnamefont {Gaeta}},\ and\ \bibinfo {author} {\bibfnamefont {M.}~\bibnamefont {Lipson}},\ }\bibfield  {title} {\bibinfo {title} {Ultra-low-loss on-chip resonators with sub-milliwatt parametric oscillation threshold},\ }\href@noop {} {\bibfield  {journal} {\bibinfo  {journal} {Optica}\ }\textbf {\bibinfo {volume} {4}},\ \bibinfo {pages} {619} (\bibinfo {year} {2017})}\BibitemShut {NoStop}%
\bibitem [{\citenamefont {Liu}\ \emph {et~al.}(2018)\citenamefont {Liu}, \citenamefont {Raja}, \citenamefont {Karpov}, \citenamefont {Ghadiani}, \citenamefont {Pfeiffer}, \citenamefont {Du}, \citenamefont {Engelsen}, \citenamefont {Guo}, \citenamefont {Zervas},\ and\ \citenamefont {Kippenberg}}]{liu2018ultralow}%
  \BibitemOpen
  \bibfield  {author} {\bibinfo {author} {\bibfnamefont {J.}~\bibnamefont {Liu}}, \bibinfo {author} {\bibfnamefont {A.~S.}\ \bibnamefont {Raja}}, \bibinfo {author} {\bibfnamefont {M.}~\bibnamefont {Karpov}}, \bibinfo {author} {\bibfnamefont {B.}~\bibnamefont {Ghadiani}}, \bibinfo {author} {\bibfnamefont {M.~H.}\ \bibnamefont {Pfeiffer}}, \bibinfo {author} {\bibfnamefont {B.}~\bibnamefont {Du}}, \bibinfo {author} {\bibfnamefont {N.~J.}\ \bibnamefont {Engelsen}}, \bibinfo {author} {\bibfnamefont {H.}~\bibnamefont {Guo}}, \bibinfo {author} {\bibfnamefont {M.}~\bibnamefont {Zervas}},\ and\ \bibinfo {author} {\bibfnamefont {T.~J.}\ \bibnamefont {Kippenberg}},\ }\bibfield  {title} {\bibinfo {title} {Ultralow-power chip-based soliton microcombs for photonic integration},\ }\href@noop {} {\bibfield  {journal} {\bibinfo  {journal} {Optica}\ }\textbf {\bibinfo {volume} {5}},\ \bibinfo {pages} {1347} (\bibinfo {year} {2018})}\BibitemShut {NoStop}%
\bibitem [{\citenamefont {Ye}\ \emph {et~al.}(2019)\citenamefont {Ye}, \citenamefont {Twayana}, \citenamefont {Andrekson},\ and\ \citenamefont {Torres-Company}}]{ye2019high}%
  \BibitemOpen
  \bibfield  {author} {\bibinfo {author} {\bibfnamefont {Z.}~\bibnamefont {Ye}}, \bibinfo {author} {\bibfnamefont {K.}~\bibnamefont {Twayana}}, \bibinfo {author} {\bibfnamefont {P.~A.}\ \bibnamefont {Andrekson}},\ and\ \bibinfo {author} {\bibfnamefont {V.}~\bibnamefont {Torres-Company}},\ }\bibfield  {title} {\bibinfo {title} {High-q si3n4 microresonators based on a subtractive processing for kerr nonlinear optics},\ }\href@noop {} {\bibfield  {journal} {\bibinfo  {journal} {Opt. Express}\ }\textbf {\bibinfo {volume} {27}},\ \bibinfo {pages} {35719} (\bibinfo {year} {2019})}\BibitemShut {NoStop}%
\bibitem [{\citenamefont {Pu}\ \emph {et~al.}(2016)\citenamefont {Pu}, \citenamefont {Ottaviano}, \citenamefont {Semenova},\ and\ \citenamefont {Yvind}}]{pu2016efficient}%
  \BibitemOpen
  \bibfield  {author} {\bibinfo {author} {\bibfnamefont {M.}~\bibnamefont {Pu}}, \bibinfo {author} {\bibfnamefont {L.}~\bibnamefont {Ottaviano}}, \bibinfo {author} {\bibfnamefont {E.}~\bibnamefont {Semenova}},\ and\ \bibinfo {author} {\bibfnamefont {K.}~\bibnamefont {Yvind}},\ }\bibfield  {title} {\bibinfo {title} {Efficient frequency comb generation in algaas-on-insulator},\ }\href@noop {} {\bibfield  {journal} {\bibinfo  {journal} {Optica}\ }\textbf {\bibinfo {volume} {3}},\ \bibinfo {pages} {823} (\bibinfo {year} {2016})}\BibitemShut {NoStop}%
\bibitem [{\citenamefont {Chang}\ \emph {et~al.}(2020)\citenamefont {Chang}, \citenamefont {Xie}, \citenamefont {Shu}, \citenamefont {Yang}, \citenamefont {Shen}, \citenamefont {Boes}, \citenamefont {Peters}, \citenamefont {Jin}, \citenamefont {Xiang}, \citenamefont {Liu} \emph {et~al.}}]{chang2020ultra}%
  \BibitemOpen
  \bibfield  {author} {\bibinfo {author} {\bibfnamefont {L.}~\bibnamefont {Chang}}, \bibinfo {author} {\bibfnamefont {W.}~\bibnamefont {Xie}}, \bibinfo {author} {\bibfnamefont {H.}~\bibnamefont {Shu}}, \bibinfo {author} {\bibfnamefont {Q.-F.}\ \bibnamefont {Yang}}, \bibinfo {author} {\bibfnamefont {B.}~\bibnamefont {Shen}}, \bibinfo {author} {\bibfnamefont {A.}~\bibnamefont {Boes}}, \bibinfo {author} {\bibfnamefont {J.~D.}\ \bibnamefont {Peters}}, \bibinfo {author} {\bibfnamefont {W.}~\bibnamefont {Jin}}, \bibinfo {author} {\bibfnamefont {C.}~\bibnamefont {Xiang}}, \bibinfo {author} {\bibfnamefont {S.}~\bibnamefont {Liu}}, \emph {et~al.},\ }\bibfield  {title} {\bibinfo {title} {Ultra-efficient frequency comb generation in algaas-on-insulator microresonators},\ }\href@noop {} {\bibfield  {journal} {\bibinfo  {journal} {Nat. Commun.}\ }\textbf {\bibinfo {volume} {11}},\ \bibinfo {pages} {1331} (\bibinfo {year} {2020})}\BibitemShut {NoStop}%
\bibitem [{\citenamefont {Jung}\ \emph {et~al.}(2013)\citenamefont {Jung}, \citenamefont {Fong}, \citenamefont {Xiong},\ and\ \citenamefont {Tang}}]{jung2013electrical}%
  \BibitemOpen
  \bibfield  {author} {\bibinfo {author} {\bibfnamefont {H.}~\bibnamefont {Jung}}, \bibinfo {author} {\bibfnamefont {K.~Y.}\ \bibnamefont {Fong}}, \bibinfo {author} {\bibfnamefont {C.}~\bibnamefont {Xiong}},\ and\ \bibinfo {author} {\bibfnamefont {H.~X.}\ \bibnamefont {Tang}},\ }\bibfield  {title} {\bibinfo {title} {Electrical tuning and switching of an optical frequency comb generated in aluminum nitride microring resonators},\ }\href@noop {} {\bibfield  {journal} {\bibinfo  {journal} {Opt. Lett.}\ }\textbf {\bibinfo {volume} {39}},\ \bibinfo {pages} {84} (\bibinfo {year} {2013})}\BibitemShut {NoStop}%
\bibitem [{\citenamefont {Guidry}\ \emph {et~al.}(2020)\citenamefont {Guidry}, \citenamefont {Yang}, \citenamefont {Lukin}, \citenamefont {Markosyan}, \citenamefont {Yang}, \citenamefont {Fejer},\ and\ \citenamefont {Vu{\v{c}}kovi{\'c}}}]{guidry2020optical}%
  \BibitemOpen
  \bibfield  {author} {\bibinfo {author} {\bibfnamefont {M.~A.}\ \bibnamefont {Guidry}}, \bibinfo {author} {\bibfnamefont {K.~Y.}\ \bibnamefont {Yang}}, \bibinfo {author} {\bibfnamefont {D.~M.}\ \bibnamefont {Lukin}}, \bibinfo {author} {\bibfnamefont {A.}~\bibnamefont {Markosyan}}, \bibinfo {author} {\bibfnamefont {J.}~\bibnamefont {Yang}}, \bibinfo {author} {\bibfnamefont {M.~M.}\ \bibnamefont {Fejer}},\ and\ \bibinfo {author} {\bibfnamefont {J.}~\bibnamefont {Vu{\v{c}}kovi{\'c}}},\ }\bibfield  {title} {\bibinfo {title} {Optical parametric oscillation in silicon carbide nanophotonics},\ }\href@noop {} {\bibfield  {journal} {\bibinfo  {journal} {Optica}\ }\textbf {\bibinfo {volume} {7}},\ \bibinfo {pages} {1139} (\bibinfo {year} {2020})}\BibitemShut {NoStop}%
\bibitem [{\citenamefont {Zhang}\ \emph {et~al.}(2017)\citenamefont {Zhang}, \citenamefont {Wang}, \citenamefont {Cheng}, \citenamefont {Shams-Ansari},\ and\ \citenamefont {Lon{\v{c}}ar}}]{zhang2017monolithic}%
  \BibitemOpen
  \bibfield  {author} {\bibinfo {author} {\bibfnamefont {M.}~\bibnamefont {Zhang}}, \bibinfo {author} {\bibfnamefont {C.}~\bibnamefont {Wang}}, \bibinfo {author} {\bibfnamefont {R.}~\bibnamefont {Cheng}}, \bibinfo {author} {\bibfnamefont {A.}~\bibnamefont {Shams-Ansari}},\ and\ \bibinfo {author} {\bibfnamefont {M.}~\bibnamefont {Lon{\v{c}}ar}},\ }\bibfield  {title} {\bibinfo {title} {Monolithic ultra-high-{Q} lithium niobate microring resonator},\ }\href@noop {} {\bibfield  {journal} {\bibinfo  {journal} {Optica}\ }\textbf {\bibinfo {volume} {4}},\ \bibinfo {pages} {1536} (\bibinfo {year} {2017})}\BibitemShut {NoStop}%
\bibitem [{\citenamefont {He}\ \emph {et~al.}(2019{\natexlab{a}})\citenamefont {He}, \citenamefont {Yang}, \citenamefont {Ling}, \citenamefont {Luo}, \citenamefont {Liang}, \citenamefont {Li}, \citenamefont {Shen}, \citenamefont {Wang}, \citenamefont {Vahala},\ and\ \citenamefont {Lin}}]{he2019self}%
  \BibitemOpen
  \bibfield  {author} {\bibinfo {author} {\bibfnamefont {Y.}~\bibnamefont {He}}, \bibinfo {author} {\bibfnamefont {Q.-F.}\ \bibnamefont {Yang}}, \bibinfo {author} {\bibfnamefont {J.}~\bibnamefont {Ling}}, \bibinfo {author} {\bibfnamefont {R.}~\bibnamefont {Luo}}, \bibinfo {author} {\bibfnamefont {H.}~\bibnamefont {Liang}}, \bibinfo {author} {\bibfnamefont {M.}~\bibnamefont {Li}}, \bibinfo {author} {\bibfnamefont {B.}~\bibnamefont {Shen}}, \bibinfo {author} {\bibfnamefont {H.}~\bibnamefont {Wang}}, \bibinfo {author} {\bibfnamefont {K.~J.}\ \bibnamefont {Vahala}},\ and\ \bibinfo {author} {\bibfnamefont {Q.}~\bibnamefont {Lin}},\ }\bibfield  {title} {\bibinfo {title} {Self-starting bi-chromatic {L}i{N}b{O}3 soliton microcomb},\ }\href@noop {} {\bibfield  {journal} {\bibinfo  {journal} {Optica}\ }\textbf {\bibinfo {volume} {6}},\ \bibinfo {pages} {1138} (\bibinfo {year} {2019}{\natexlab{a}})}\BibitemShut {NoStop}%
\bibitem [{\citenamefont {Gong}\ \emph {et~al.}(2019)\citenamefont {Gong}, \citenamefont {Liu}, \citenamefont {Xu}, \citenamefont {Xu}, \citenamefont {Surya}, \citenamefont {Lu}, \citenamefont {Bruch}, \citenamefont {Zou},\ and\ \citenamefont {Tang}}]{gong2019soliton}%
  \BibitemOpen
  \bibfield  {author} {\bibinfo {author} {\bibfnamefont {Z.}~\bibnamefont {Gong}}, \bibinfo {author} {\bibfnamefont {X.}~\bibnamefont {Liu}}, \bibinfo {author} {\bibfnamefont {Y.}~\bibnamefont {Xu}}, \bibinfo {author} {\bibfnamefont {M.}~\bibnamefont {Xu}}, \bibinfo {author} {\bibfnamefont {J.~B.}\ \bibnamefont {Surya}}, \bibinfo {author} {\bibfnamefont {J.}~\bibnamefont {Lu}}, \bibinfo {author} {\bibfnamefont {A.}~\bibnamefont {Bruch}}, \bibinfo {author} {\bibfnamefont {C.}~\bibnamefont {Zou}},\ and\ \bibinfo {author} {\bibfnamefont {H.~X.}\ \bibnamefont {Tang}},\ }\bibfield  {title} {\bibinfo {title} {Soliton microcomb generation at 2 $\mu$m in z-cut lithium niobate microring resonators},\ }\href@noop {} {\bibfield  {journal} {\bibinfo  {journal} {Opt. Lett.}\ }\textbf {\bibinfo {volume} {44}},\ \bibinfo {pages} {3182} (\bibinfo {year} {2019})}\BibitemShut {NoStop}%
\bibitem [{\citenamefont {Gao}\ \emph {et~al.}(2023)\citenamefont {Gao}, \citenamefont {Lei}, \citenamefont {Girardi}, \citenamefont {Ye}, \citenamefont {Van~Laer}, \citenamefont {Torres-Company},\ and\ \citenamefont {Schr{\"o}der}}]{gao2023compact}%
  \BibitemOpen
  \bibfield  {author} {\bibinfo {author} {\bibfnamefont {Y.}~\bibnamefont {Gao}}, \bibinfo {author} {\bibfnamefont {F.}~\bibnamefont {Lei}}, \bibinfo {author} {\bibfnamefont {M.}~\bibnamefont {Girardi}}, \bibinfo {author} {\bibfnamefont {Z.}~\bibnamefont {Ye}}, \bibinfo {author} {\bibfnamefont {R.}~\bibnamefont {Van~Laer}}, \bibinfo {author} {\bibfnamefont {V.}~\bibnamefont {Torres-Company}},\ and\ \bibinfo {author} {\bibfnamefont {J.}~\bibnamefont {Schr{\"o}der}},\ }\bibfield  {title} {\bibinfo {title} {Compact lithium niobate microring resonators in the ultrahigh q/v regime},\ }\href@noop {} {\bibfield  {journal} {\bibinfo  {journal} {Opt. Lett.}\ }\textbf {\bibinfo {volume} {48}},\ \bibinfo {pages} {3949} (\bibinfo {year} {2023})}\BibitemShut {NoStop}%
\bibitem [{\citenamefont {Zhu}\ \emph {et~al.}(2021)\citenamefont {Zhu}, \citenamefont {Shao}, \citenamefont {Yu}, \citenamefont {Cheng}, \citenamefont {Desiatov}, \citenamefont {Xin}, \citenamefont {Hu}, \citenamefont {Holzgrafe}, \citenamefont {Ghosh}, \citenamefont {Shams-Ansari}, \citenamefont {Puma}, \citenamefont {Sinclair}, \citenamefont {Reimer}, \citenamefont {Zhang},\ and\ \citenamefont {Lončar}}]{zhu2021integrated}%
  \BibitemOpen
  \bibfield  {author} {\bibinfo {author} {\bibfnamefont {D.}~\bibnamefont {Zhu}}, \bibinfo {author} {\bibfnamefont {L.}~\bibnamefont {Shao}}, \bibinfo {author} {\bibfnamefont {M.}~\bibnamefont {Yu}}, \bibinfo {author} {\bibfnamefont {R.}~\bibnamefont {Cheng}}, \bibinfo {author} {\bibfnamefont {B.}~\bibnamefont {Desiatov}}, \bibinfo {author} {\bibfnamefont {C.}~\bibnamefont {Xin}}, \bibinfo {author} {\bibfnamefont {Y.}~\bibnamefont {Hu}}, \bibinfo {author} {\bibfnamefont {J.}~\bibnamefont {Holzgrafe}}, \bibinfo {author} {\bibfnamefont {S.}~\bibnamefont {Ghosh}}, \bibinfo {author} {\bibfnamefont {A.}~\bibnamefont {Shams-Ansari}}, \bibinfo {author} {\bibfnamefont {E.}~\bibnamefont {Puma}}, \bibinfo {author} {\bibfnamefont {N.}~\bibnamefont {Sinclair}}, \bibinfo {author} {\bibfnamefont {C.}~\bibnamefont {Reimer}}, \bibinfo {author} {\bibfnamefont {M.}~\bibnamefont {Zhang}},\ and\ \bibinfo {author} {\bibfnamefont {M.}~\bibnamefont {Lončar}},\ }\bibfield  {title} {\bibinfo {title} {Integrated photonics on
  thin-film lithium niobate},\ }\href@noop {} {\bibfield  {journal} {\bibinfo  {journal} {Adv. Opt. Photonics}\ }\textbf {\bibinfo {volume} {13}},\ \bibinfo {pages} {242} (\bibinfo {year} {2021})}\BibitemShut {NoStop}%
\bibitem [{\citenamefont {Qi}\ and\ \citenamefont {Li}(2020)}]{qi2020integrated}%
  \BibitemOpen
  \bibfield  {author} {\bibinfo {author} {\bibfnamefont {Y.}~\bibnamefont {Qi}}\ and\ \bibinfo {author} {\bibfnamefont {Y.}~\bibnamefont {Li}},\ }\bibfield  {title} {\bibinfo {title} {Integrated lithium niobate photonics},\ }\href@noop {} {\bibfield  {journal} {\bibinfo  {journal} {Nanophotonics}\ }\textbf {\bibinfo {volume} {9}},\ \bibinfo {pages} {1287} (\bibinfo {year} {2020})}\BibitemShut {NoStop}%
\bibitem [{\citenamefont {Honardoost}\ \emph {et~al.}(2020)\citenamefont {Honardoost}, \citenamefont {Abdelsalam},\ and\ \citenamefont {Fathpour}}]{honardoost2020rejuvenating}%
  \BibitemOpen
  \bibfield  {author} {\bibinfo {author} {\bibfnamefont {A.}~\bibnamefont {Honardoost}}, \bibinfo {author} {\bibfnamefont {K.}~\bibnamefont {Abdelsalam}},\ and\ \bibinfo {author} {\bibfnamefont {S.}~\bibnamefont {Fathpour}},\ }\bibfield  {title} {\bibinfo {title} {Rejuvenating a versatile photonic material: thin-film lithium niobate},\ }\href@noop {} {\bibfield  {journal} {\bibinfo  {journal} {Laser Photonics Rev.}\ }\textbf {\bibinfo {volume} {14}},\ \bibinfo {pages} {2000088} (\bibinfo {year} {2020})}\BibitemShut {NoStop}%
\bibitem [{\citenamefont {Weis}\ and\ \citenamefont {Gaylord}(1985)}]{weis1985lithium}%
  \BibitemOpen
  \bibfield  {author} {\bibinfo {author} {\bibfnamefont {R.}~\bibnamefont {Weis}}\ and\ \bibinfo {author} {\bibfnamefont {T.}~\bibnamefont {Gaylord}},\ }\bibfield  {title} {\bibinfo {title} {Lithium niobate: {S}ummary of physical properties and crystal structure},\ }\href@noop {} {\bibfield  {journal} {\bibinfo  {journal} {Appl. Phys. A}\ }\textbf {\bibinfo {volume} {37}},\ \bibinfo {pages} {191} (\bibinfo {year} {1985})}\BibitemShut {NoStop}%
\bibitem [{\citenamefont {Wang}\ \emph {et~al.}(2018)\citenamefont {Wang}, \citenamefont {Zhang}, \citenamefont {Chen}, \citenamefont {Bertrand}, \citenamefont {Shams-Ansari}, \citenamefont {Chandrasekhar}, \citenamefont {Winzer},\ and\ \citenamefont {Lon{\v{c}}ar}}]{wang2018integrated}%
  \BibitemOpen
  \bibfield  {author} {\bibinfo {author} {\bibfnamefont {C.}~\bibnamefont {Wang}}, \bibinfo {author} {\bibfnamefont {M.}~\bibnamefont {Zhang}}, \bibinfo {author} {\bibfnamefont {X.}~\bibnamefont {Chen}}, \bibinfo {author} {\bibfnamefont {M.}~\bibnamefont {Bertrand}}, \bibinfo {author} {\bibfnamefont {A.}~\bibnamefont {Shams-Ansari}}, \bibinfo {author} {\bibfnamefont {S.}~\bibnamefont {Chandrasekhar}}, \bibinfo {author} {\bibfnamefont {P.}~\bibnamefont {Winzer}},\ and\ \bibinfo {author} {\bibfnamefont {M.}~\bibnamefont {Lon{\v{c}}ar}},\ }\bibfield  {title} {\bibinfo {title} {Integrated lithium niobate electro-optic modulators operating at {CMOS}-compatible voltages},\ }\href@noop {} {\bibfield  {journal} {\bibinfo  {journal} {Nature}\ }\textbf {\bibinfo {volume} {562}},\ \bibinfo {pages} {101} (\bibinfo {year} {2018})}\BibitemShut {NoStop}%
\bibitem [{\citenamefont {He}\ \emph {et~al.}(2019{\natexlab{b}})\citenamefont {He}, \citenamefont {Xu}, \citenamefont {Ren}, \citenamefont {Jian}, \citenamefont {Ruan}, \citenamefont {Xu}, \citenamefont {Gao}, \citenamefont {Sun}, \citenamefont {Wen}, \citenamefont {Zhou}, \citenamefont {Liu}, \citenamefont {Guo}, \citenamefont {Chen}, \citenamefont {Yu}, \citenamefont {Liu},\ and\ \citenamefont {Cai}}]{he2019high}%
  \BibitemOpen
  \bibfield  {author} {\bibinfo {author} {\bibfnamefont {M.}~\bibnamefont {He}}, \bibinfo {author} {\bibfnamefont {M.}~\bibnamefont {Xu}}, \bibinfo {author} {\bibfnamefont {Y.}~\bibnamefont {Ren}}, \bibinfo {author} {\bibfnamefont {J.}~\bibnamefont {Jian}}, \bibinfo {author} {\bibfnamefont {Z.}~\bibnamefont {Ruan}}, \bibinfo {author} {\bibfnamefont {Y.}~\bibnamefont {Xu}}, \bibinfo {author} {\bibfnamefont {S.}~\bibnamefont {Gao}}, \bibinfo {author} {\bibfnamefont {S.}~\bibnamefont {Sun}}, \bibinfo {author} {\bibfnamefont {X.}~\bibnamefont {Wen}}, \bibinfo {author} {\bibfnamefont {L.}~\bibnamefont {Zhou}}, \bibinfo {author} {\bibfnamefont {L.}~\bibnamefont {Liu}}, \bibinfo {author} {\bibfnamefont {C.}~\bibnamefont {Guo}}, \bibinfo {author} {\bibfnamefont {H.}~\bibnamefont {Chen}}, \bibinfo {author} {\bibfnamefont {S.}~\bibnamefont {Yu}}, \bibinfo {author} {\bibfnamefont {L.}~\bibnamefont {Liu}},\ and\ \bibinfo {author} {\bibfnamefont {X.}~\bibnamefont {Cai}},\ }\bibfield  {title} {\bibinfo {title}
  {High-performance hybrid silicon and lithium niobate {M}ach--{Z}ehnder modulators for 100 gbit s- 1 and beyond},\ }\href@noop {} {\bibfield  {journal} {\bibinfo  {journal} {Nat. Photonics}\ }\textbf {\bibinfo {volume} {13}},\ \bibinfo {pages} {359} (\bibinfo {year} {2019}{\natexlab{b}})}\BibitemShut {NoStop}%
\bibitem [{\citenamefont {Zhang}\ \emph {et~al.}(2019)\citenamefont {Zhang}, \citenamefont {Buscaino}, \citenamefont {Wang}, \citenamefont {Shams-Ansari}, \citenamefont {Reimer}, \citenamefont {Zhu}, \citenamefont {Kahn},\ and\ \citenamefont {Lon{\v{c}}ar}}]{zhang2019broadband}%
  \BibitemOpen
  \bibfield  {author} {\bibinfo {author} {\bibfnamefont {M.}~\bibnamefont {Zhang}}, \bibinfo {author} {\bibfnamefont {B.}~\bibnamefont {Buscaino}}, \bibinfo {author} {\bibfnamefont {C.}~\bibnamefont {Wang}}, \bibinfo {author} {\bibfnamefont {A.}~\bibnamefont {Shams-Ansari}}, \bibinfo {author} {\bibfnamefont {C.}~\bibnamefont {Reimer}}, \bibinfo {author} {\bibfnamefont {R.}~\bibnamefont {Zhu}}, \bibinfo {author} {\bibfnamefont {J.~M.}\ \bibnamefont {Kahn}},\ and\ \bibinfo {author} {\bibfnamefont {M.}~\bibnamefont {Lon{\v{c}}ar}},\ }\bibfield  {title} {\bibinfo {title} {Broadband electro-optic frequency comb generation in a lithium niobate microring resonator},\ }\href@noop {} {\bibfield  {journal} {\bibinfo  {journal} {Nature}\ }\textbf {\bibinfo {volume} {568}},\ \bibinfo {pages} {373} (\bibinfo {year} {2019})}\BibitemShut {NoStop}%
\bibitem [{\citenamefont {Lu}\ \emph {et~al.}(2019)\citenamefont {Lu}, \citenamefont {Surya}, \citenamefont {Liu}, \citenamefont {Bruch}, \citenamefont {Gong}, \citenamefont {Xu},\ and\ \citenamefont {Tang}}]{lu2019periodically}%
  \BibitemOpen
  \bibfield  {author} {\bibinfo {author} {\bibfnamefont {J.}~\bibnamefont {Lu}}, \bibinfo {author} {\bibfnamefont {J.~B.}\ \bibnamefont {Surya}}, \bibinfo {author} {\bibfnamefont {X.}~\bibnamefont {Liu}}, \bibinfo {author} {\bibfnamefont {A.~W.}\ \bibnamefont {Bruch}}, \bibinfo {author} {\bibfnamefont {Z.}~\bibnamefont {Gong}}, \bibinfo {author} {\bibfnamefont {Y.}~\bibnamefont {Xu}},\ and\ \bibinfo {author} {\bibfnamefont {H.~X.}\ \bibnamefont {Tang}},\ }\bibfield  {title} {\bibinfo {title} {Periodically poled thin-film lithium niobate microring resonators with a second-harmonic generation efficiency of 250,000\%/{W}},\ }\href@noop {} {\bibfield  {journal} {\bibinfo  {journal} {Optica}\ }\textbf {\bibinfo {volume} {6}},\ \bibinfo {pages} {1455} (\bibinfo {year} {2019})}\BibitemShut {NoStop}%
\bibitem [{\citenamefont {Chen}\ \emph {et~al.}(2019)\citenamefont {Chen}, \citenamefont {Ma}, \citenamefont {Sua}, \citenamefont {Li}, \citenamefont {Tang},\ and\ \citenamefont {Huang}}]{chen2019ultra}%
  \BibitemOpen
  \bibfield  {author} {\bibinfo {author} {\bibfnamefont {J.-Y.}\ \bibnamefont {Chen}}, \bibinfo {author} {\bibfnamefont {Z.-H.}\ \bibnamefont {Ma}}, \bibinfo {author} {\bibfnamefont {Y.~M.}\ \bibnamefont {Sua}}, \bibinfo {author} {\bibfnamefont {Z.}~\bibnamefont {Li}}, \bibinfo {author} {\bibfnamefont {C.}~\bibnamefont {Tang}},\ and\ \bibinfo {author} {\bibfnamefont {Y.-P.}\ \bibnamefont {Huang}},\ }\bibfield  {title} {\bibinfo {title} {Ultra-efficient frequency conversion in quasi-phase-matched lithium niobate microrings},\ }\href@noop {} {\bibfield  {journal} {\bibinfo  {journal} {Optica}\ }\textbf {\bibinfo {volume} {6}},\ \bibinfo {pages} {1244} (\bibinfo {year} {2019})}\BibitemShut {NoStop}%
\bibitem [{\citenamefont {Nehra}\ \emph {et~al.}(2022)\citenamefont {Nehra}, \citenamefont {Sekine}, \citenamefont {Ledezma}, \citenamefont {Guo}, \citenamefont {Gray}, \citenamefont {Roy},\ and\ \citenamefont {Marandi}}]{nehra2022few}%
  \BibitemOpen
  \bibfield  {author} {\bibinfo {author} {\bibfnamefont {R.}~\bibnamefont {Nehra}}, \bibinfo {author} {\bibfnamefont {R.}~\bibnamefont {Sekine}}, \bibinfo {author} {\bibfnamefont {L.}~\bibnamefont {Ledezma}}, \bibinfo {author} {\bibfnamefont {Q.}~\bibnamefont {Guo}}, \bibinfo {author} {\bibfnamefont {R.~M.}\ \bibnamefont {Gray}}, \bibinfo {author} {\bibfnamefont {A.}~\bibnamefont {Roy}},\ and\ \bibinfo {author} {\bibfnamefont {A.}~\bibnamefont {Marandi}},\ }\bibfield  {title} {\bibinfo {title} {Few-cycle vacuum squeezing in nanophotonics},\ }\href@noop {} {\bibfield  {journal} {\bibinfo  {journal} {arXiv preprint arXiv:2201.06768}\ } (\bibinfo {year} {2022})}\BibitemShut {NoStop}%
\bibitem [{\citenamefont {Zhao}\ \emph {et~al.}(2020)\citenamefont {Zhao}, \citenamefont {Ma}, \citenamefont {R{\"u}sing},\ and\ \citenamefont {Mookherjea}}]{zhao2020high}%
  \BibitemOpen
  \bibfield  {author} {\bibinfo {author} {\bibfnamefont {J.}~\bibnamefont {Zhao}}, \bibinfo {author} {\bibfnamefont {C.}~\bibnamefont {Ma}}, \bibinfo {author} {\bibfnamefont {M.}~\bibnamefont {R{\"u}sing}},\ and\ \bibinfo {author} {\bibfnamefont {S.}~\bibnamefont {Mookherjea}},\ }\bibfield  {title} {\bibinfo {title} {High quality entangled photon pair generation in periodically poled thin-film lithium niobate waveguides},\ }\href@noop {} {\bibfield  {journal} {\bibinfo  {journal} {Phys. Rev. Lett.}\ }\textbf {\bibinfo {volume} {124}},\ \bibinfo {pages} {163603} (\bibinfo {year} {2020})}\BibitemShut {NoStop}%
\bibitem [{\citenamefont {Lu}\ \emph {et~al.}(2021)\citenamefont {Lu}, \citenamefont {Al~Sayem}, \citenamefont {Gong}, \citenamefont {Surya}, \citenamefont {Zou},\ and\ \citenamefont {Tang}}]{lu2021ultralow}%
  \BibitemOpen
  \bibfield  {author} {\bibinfo {author} {\bibfnamefont {J.}~\bibnamefont {Lu}}, \bibinfo {author} {\bibfnamefont {A.}~\bibnamefont {Al~Sayem}}, \bibinfo {author} {\bibfnamefont {Z.}~\bibnamefont {Gong}}, \bibinfo {author} {\bibfnamefont {J.~B.}\ \bibnamefont {Surya}}, \bibinfo {author} {\bibfnamefont {C.-L.}\ \bibnamefont {Zou}},\ and\ \bibinfo {author} {\bibfnamefont {H.~X.}\ \bibnamefont {Tang}},\ }\bibfield  {title} {\bibinfo {title} {Ultralow-threshold thin-film lithium niobate optical parametric oscillator},\ }\href@noop {} {\bibfield  {journal} {\bibinfo  {journal} {Optica}\ }\textbf {\bibinfo {volume} {8}},\ \bibinfo {pages} {539} (\bibinfo {year} {2021})}\BibitemShut {NoStop}%
\bibitem [{\citenamefont {McKenna}\ \emph {et~al.}(2022)\citenamefont {McKenna}, \citenamefont {Stokowski}, \citenamefont {Ansari}, \citenamefont {Mishra}, \citenamefont {Jankowski}, \citenamefont {Sarabalis}, \citenamefont {Herrmann}, \citenamefont {Langrock}, \citenamefont {Fejer},\ and\ \citenamefont {Safavi-Naeini}}]{mckenna2022ultra}%
  \BibitemOpen
  \bibfield  {author} {\bibinfo {author} {\bibfnamefont {T.~P.}\ \bibnamefont {McKenna}}, \bibinfo {author} {\bibfnamefont {H.~S.}\ \bibnamefont {Stokowski}}, \bibinfo {author} {\bibfnamefont {V.}~\bibnamefont {Ansari}}, \bibinfo {author} {\bibfnamefont {J.}~\bibnamefont {Mishra}}, \bibinfo {author} {\bibfnamefont {M.}~\bibnamefont {Jankowski}}, \bibinfo {author} {\bibfnamefont {C.~J.}\ \bibnamefont {Sarabalis}}, \bibinfo {author} {\bibfnamefont {J.~F.}\ \bibnamefont {Herrmann}}, \bibinfo {author} {\bibfnamefont {C.}~\bibnamefont {Langrock}}, \bibinfo {author} {\bibfnamefont {M.~M.}\ \bibnamefont {Fejer}},\ and\ \bibinfo {author} {\bibfnamefont {A.~H.}\ \bibnamefont {Safavi-Naeini}},\ }\bibfield  {title} {\bibinfo {title} {Ultra-low-power second-order nonlinear optics on a chip},\ }\href@noop {} {\bibfield  {journal} {\bibinfo  {journal} {Nat. Commun.}\ }\textbf {\bibinfo {volume} {13}},\ \bibinfo {pages} {4532} (\bibinfo {year} {2022})}\BibitemShut {NoStop}%
\bibitem [{\citenamefont {Xiang}\ \emph {et~al.}(2020)\citenamefont {Xiang}, \citenamefont {Jin}, \citenamefont {Guo}, \citenamefont {Peters}, \citenamefont {Kennedy}, \citenamefont {Selvidge}, \citenamefont {Morton},\ and\ \citenamefont {Bowers}}]{xiang2020narrow}%
  \BibitemOpen
  \bibfield  {author} {\bibinfo {author} {\bibfnamefont {C.}~\bibnamefont {Xiang}}, \bibinfo {author} {\bibfnamefont {W.}~\bibnamefont {Jin}}, \bibinfo {author} {\bibfnamefont {J.}~\bibnamefont {Guo}}, \bibinfo {author} {\bibfnamefont {J.~D.}\ \bibnamefont {Peters}}, \bibinfo {author} {\bibfnamefont {M.}~\bibnamefont {Kennedy}}, \bibinfo {author} {\bibfnamefont {J.}~\bibnamefont {Selvidge}}, \bibinfo {author} {\bibfnamefont {P.~A.}\ \bibnamefont {Morton}},\ and\ \bibinfo {author} {\bibfnamefont {J.~E.}\ \bibnamefont {Bowers}},\ }\bibfield  {title} {\bibinfo {title} {Narrow-linewidth iii-v/si/si 3 n 4 laser using multilayer heterogeneous integration},\ }\href@noop {} {\bibfield  {journal} {\bibinfo  {journal} {Optica}\ }\textbf {\bibinfo {volume} {7}},\ \bibinfo {pages} {20} (\bibinfo {year} {2020})}\BibitemShut {NoStop}%
\bibitem [{\citenamefont {Lee}\ \emph {et~al.}(2012)\citenamefont {Lee}, \citenamefont {Chen}, \citenamefont {Li}, \citenamefont {Painter},\ and\ \citenamefont {Vahala}}]{lee2012ultra}%
  \BibitemOpen
  \bibfield  {author} {\bibinfo {author} {\bibfnamefont {H.}~\bibnamefont {Lee}}, \bibinfo {author} {\bibfnamefont {T.}~\bibnamefont {Chen}}, \bibinfo {author} {\bibfnamefont {J.}~\bibnamefont {Li}}, \bibinfo {author} {\bibfnamefont {O.}~\bibnamefont {Painter}},\ and\ \bibinfo {author} {\bibfnamefont {K.~J.}\ \bibnamefont {Vahala}},\ }\bibfield  {title} {\bibinfo {title} {Ultra-low-loss optical delay line on a silicon chip},\ }\href@noop {} {\bibfield  {journal} {\bibinfo  {journal} {Nat. Commun.}\ }\textbf {\bibinfo {volume} {3}},\ \bibinfo {pages} {867} (\bibinfo {year} {2012})}\BibitemShut {NoStop}%
\bibitem [{\citenamefont {Hong}\ \emph {et~al.}(2021)\citenamefont {Hong}, \citenamefont {Zhang}, \citenamefont {Wang}, \citenamefont {Zhang}, \citenamefont {Xie},\ and\ \citenamefont {Dai}}]{hong2021ultralow}%
  \BibitemOpen
  \bibfield  {author} {\bibinfo {author} {\bibfnamefont {S.}~\bibnamefont {Hong}}, \bibinfo {author} {\bibfnamefont {L.}~\bibnamefont {Zhang}}, \bibinfo {author} {\bibfnamefont {Y.}~\bibnamefont {Wang}}, \bibinfo {author} {\bibfnamefont {M.}~\bibnamefont {Zhang}}, \bibinfo {author} {\bibfnamefont {Y.}~\bibnamefont {Xie}},\ and\ \bibinfo {author} {\bibfnamefont {D.}~\bibnamefont {Dai}},\ }\bibfield  {title} {\bibinfo {title} {Ultralow-loss compact silicon photonic waveguide spirals and delay lines},\ }\href@noop {} {\bibfield  {journal} {\bibinfo  {journal} {Photonics Res.}\ }\textbf {\bibinfo {volume} {10}},\ \bibinfo {pages} {1} (\bibinfo {year} {2021})}\BibitemShut {NoStop}%
\bibitem [{\citenamefont {Ji}\ \emph {et~al.}(2019)\citenamefont {Ji}, \citenamefont {Yao}, \citenamefont {Gan}, \citenamefont {Mohanty}, \citenamefont {Tadayon}, \citenamefont {Hendon},\ and\ \citenamefont {Lipson}}]{ji2019chip}%
  \BibitemOpen
  \bibfield  {author} {\bibinfo {author} {\bibfnamefont {X.}~\bibnamefont {Ji}}, \bibinfo {author} {\bibfnamefont {X.}~\bibnamefont {Yao}}, \bibinfo {author} {\bibfnamefont {Y.}~\bibnamefont {Gan}}, \bibinfo {author} {\bibfnamefont {A.}~\bibnamefont {Mohanty}}, \bibinfo {author} {\bibfnamefont {M.~A.}\ \bibnamefont {Tadayon}}, \bibinfo {author} {\bibfnamefont {C.~P.}\ \bibnamefont {Hendon}},\ and\ \bibinfo {author} {\bibfnamefont {M.}~\bibnamefont {Lipson}},\ }\bibfield  {title} {\bibinfo {title} {On-chip tunable photonic delay line},\ }\href@noop {} {\bibfield  {journal} {\bibinfo  {journal} {APL Photonics}\ }\textbf {\bibinfo {volume} {4}} (\bibinfo {year} {2019})}\BibitemShut {NoStop}%
\bibitem [{\citenamefont {Ye}\ \emph {et~al.}(2021)\citenamefont {Ye}, \citenamefont {Zhao}, \citenamefont {Twayana}, \citenamefont {Karlsson}, \citenamefont {Torres-Company},\ and\ \citenamefont {Andrekson}}]{ye2021overcoming}%
  \BibitemOpen
  \bibfield  {author} {\bibinfo {author} {\bibfnamefont {Z.}~\bibnamefont {Ye}}, \bibinfo {author} {\bibfnamefont {P.}~\bibnamefont {Zhao}}, \bibinfo {author} {\bibfnamefont {K.}~\bibnamefont {Twayana}}, \bibinfo {author} {\bibfnamefont {M.}~\bibnamefont {Karlsson}}, \bibinfo {author} {\bibfnamefont {V.}~\bibnamefont {Torres-Company}},\ and\ \bibinfo {author} {\bibfnamefont {P.~A.}\ \bibnamefont {Andrekson}},\ }\bibfield  {title} {\bibinfo {title} {Overcoming the quantum limit of optical amplification in monolithic waveguides},\ }\href@noop {} {\bibfield  {journal} {\bibinfo  {journal} {Sci. Adv.}\ }\textbf {\bibinfo {volume} {7}},\ \bibinfo {pages} {eabi8150} (\bibinfo {year} {2021})}\BibitemShut {NoStop}%
\bibitem [{\citenamefont {Kazama}\ \emph {et~al.}(2021)\citenamefont {Kazama}, \citenamefont {Umeki}, \citenamefont {Shimizu}, \citenamefont {Kashiwazaki}, \citenamefont {Enbutsu}, \citenamefont {Kasahara}, \citenamefont {Miyamoto},\ and\ \citenamefont {Watanabe}}]{kazama2021over}%
  \BibitemOpen
  \bibfield  {author} {\bibinfo {author} {\bibfnamefont {T.}~\bibnamefont {Kazama}}, \bibinfo {author} {\bibfnamefont {T.}~\bibnamefont {Umeki}}, \bibinfo {author} {\bibfnamefont {S.}~\bibnamefont {Shimizu}}, \bibinfo {author} {\bibfnamefont {T.}~\bibnamefont {Kashiwazaki}}, \bibinfo {author} {\bibfnamefont {K.}~\bibnamefont {Enbutsu}}, \bibinfo {author} {\bibfnamefont {R.}~\bibnamefont {Kasahara}}, \bibinfo {author} {\bibfnamefont {Y.}~\bibnamefont {Miyamoto}},\ and\ \bibinfo {author} {\bibfnamefont {K.}~\bibnamefont {Watanabe}},\ }\bibfield  {title} {\bibinfo {title} {Over-30-db gain and 1-db noise figure phase-sensitive amplification using a pump-combiner-integrated fiber i/o ppln module},\ }\href@noop {} {\bibfield  {journal} {\bibinfo  {journal} {Opt. Express}\ }\textbf {\bibinfo {volume} {29}},\ \bibinfo {pages} {28824} (\bibinfo {year} {2021})}\BibitemShut {NoStop}%
\bibitem [{\citenamefont {Bauters}\ \emph {et~al.}(2011)\citenamefont {Bauters}, \citenamefont {Heck}, \citenamefont {John}, \citenamefont {Dai}, \citenamefont {Tien}, \citenamefont {Barton}, \citenamefont {Leinse}, \citenamefont {Heideman}, \citenamefont {Blumenthal},\ and\ \citenamefont {Bowers}}]{bauters2011ultra}%
  \BibitemOpen
  \bibfield  {author} {\bibinfo {author} {\bibfnamefont {J.~F.}\ \bibnamefont {Bauters}}, \bibinfo {author} {\bibfnamefont {M.~J.}\ \bibnamefont {Heck}}, \bibinfo {author} {\bibfnamefont {D.}~\bibnamefont {John}}, \bibinfo {author} {\bibfnamefont {D.}~\bibnamefont {Dai}}, \bibinfo {author} {\bibfnamefont {M.-C.}\ \bibnamefont {Tien}}, \bibinfo {author} {\bibfnamefont {J.~S.}\ \bibnamefont {Barton}}, \bibinfo {author} {\bibfnamefont {A.}~\bibnamefont {Leinse}}, \bibinfo {author} {\bibfnamefont {R.~G.}\ \bibnamefont {Heideman}}, \bibinfo {author} {\bibfnamefont {D.~J.}\ \bibnamefont {Blumenthal}},\ and\ \bibinfo {author} {\bibfnamefont {J.~E.}\ \bibnamefont {Bowers}},\ }\bibfield  {title} {\bibinfo {title} {Ultra-low-loss high-aspect-ratio si 3 n 4 waveguides},\ }\href@noop {} {\bibfield  {journal} {\bibinfo  {journal} {Optics express}\ }\textbf {\bibinfo {volume} {19}},\ \bibinfo {pages} {3163} (\bibinfo {year} {2011})}\BibitemShut {NoStop}%
\bibitem [{\citenamefont {Liu}\ \emph {et~al.}(2021)\citenamefont {Liu}, \citenamefont {Huang}, \citenamefont {Wang}, \citenamefont {He}, \citenamefont {Raja}, \citenamefont {Liu}, \citenamefont {Engelsen},\ and\ \citenamefont {Kippenberg}}]{liu2021high}%
  \BibitemOpen
  \bibfield  {author} {\bibinfo {author} {\bibfnamefont {J.}~\bibnamefont {Liu}}, \bibinfo {author} {\bibfnamefont {G.}~\bibnamefont {Huang}}, \bibinfo {author} {\bibfnamefont {R.~N.}\ \bibnamefont {Wang}}, \bibinfo {author} {\bibfnamefont {J.}~\bibnamefont {He}}, \bibinfo {author} {\bibfnamefont {A.~S.}\ \bibnamefont {Raja}}, \bibinfo {author} {\bibfnamefont {T.}~\bibnamefont {Liu}}, \bibinfo {author} {\bibfnamefont {N.~J.}\ \bibnamefont {Engelsen}},\ and\ \bibinfo {author} {\bibfnamefont {T.~J.}\ \bibnamefont {Kippenberg}},\ }\bibfield  {title} {\bibinfo {title} {High-yield, wafer-scale fabrication of ultralow-loss, dispersion-engineered silicon nitride photonic circuits},\ }\href@noop {} {\bibfield  {journal} {\bibinfo  {journal} {Nat. Commun.}\ }\textbf {\bibinfo {volume} {12}},\ \bibinfo {pages} {2236} (\bibinfo {year} {2021})}\BibitemShut {NoStop}%
\bibitem [{\citenamefont {Pan}\ \emph {et~al.}(2020)\citenamefont {Pan}, \citenamefont {Tan}, \citenamefont {Chen}, \citenamefont {Liu}, \citenamefont {Shi},\ and\ \citenamefont {Dai}}]{pan2020compact}%
  \BibitemOpen
  \bibfield  {author} {\bibinfo {author} {\bibfnamefont {B.}~\bibnamefont {Pan}}, \bibinfo {author} {\bibfnamefont {Y.}~\bibnamefont {Tan}}, \bibinfo {author} {\bibfnamefont {P.}~\bibnamefont {Chen}}, \bibinfo {author} {\bibfnamefont {L.}~\bibnamefont {Liu}}, \bibinfo {author} {\bibfnamefont {Y.}~\bibnamefont {Shi}},\ and\ \bibinfo {author} {\bibfnamefont {D.}~\bibnamefont {Dai}},\ }\bibfield  {title} {\bibinfo {title} {Compact racetrack resonator on linbo 3},\ }\href@noop {} {\bibfield  {journal} {\bibinfo  {journal} {J. Light. Technol.}\ }\textbf {\bibinfo {volume} {39}},\ \bibinfo {pages} {1770} (\bibinfo {year} {2020})}\BibitemShut {NoStop}%
\bibitem [{\citenamefont {Yi}\ \emph {et~al.}(2024)\citenamefont {Yi}, \citenamefont {Guo}, \citenamefont {Ruan}, \citenamefont {Chen}, \citenamefont {Wei}, \citenamefont {Lu}, \citenamefont {Gong}, \citenamefont {Pan}, \citenamefont {Shen}, \citenamefont {Guan} \emph {et~al.}}]{yi2024anisotropy}%
  \BibitemOpen
  \bibfield  {author} {\bibinfo {author} {\bibfnamefont {J.}~\bibnamefont {Yi}}, \bibinfo {author} {\bibfnamefont {C.}~\bibnamefont {Guo}}, \bibinfo {author} {\bibfnamefont {Z.}~\bibnamefont {Ruan}}, \bibinfo {author} {\bibfnamefont {G.}~\bibnamefont {Chen}}, \bibinfo {author} {\bibfnamefont {H.}~\bibnamefont {Wei}}, \bibinfo {author} {\bibfnamefont {L.}~\bibnamefont {Lu}}, \bibinfo {author} {\bibfnamefont {S.}~\bibnamefont {Gong}}, \bibinfo {author} {\bibfnamefont {X.}~\bibnamefont {Pan}}, \bibinfo {author} {\bibfnamefont {X.}~\bibnamefont {Shen}}, \bibinfo {author} {\bibfnamefont {X.}~\bibnamefont {Guan}}, \emph {et~al.},\ }\bibfield  {title} {\bibinfo {title} {Anisotropy-free arrayed waveguide gratings on x-cut thin film lithium niobate platform of in-plane anisotropy},\ }\href@noop {} {\bibfield  {journal} {\bibinfo  {journal} {Light Sci. Appl.}\ }\textbf {\bibinfo {volume} {13}},\ \bibinfo {pages} {147} (\bibinfo {year} {2024})}\BibitemShut {NoStop}%
\bibitem [{\citenamefont {Chen}\ \emph {et~al.}(2012)\citenamefont {Chen}, \citenamefont {Lee}, \citenamefont {Li},\ and\ \citenamefont {Vahala}}]{chen2012general}%
  \BibitemOpen
  \bibfield  {author} {\bibinfo {author} {\bibfnamefont {T.}~\bibnamefont {Chen}}, \bibinfo {author} {\bibfnamefont {H.}~\bibnamefont {Lee}}, \bibinfo {author} {\bibfnamefont {J.}~\bibnamefont {Li}},\ and\ \bibinfo {author} {\bibfnamefont {K.~J.}\ \bibnamefont {Vahala}},\ }\bibfield  {title} {\bibinfo {title} {A general design algorithm for low optical loss adiabatic connections in waveguides},\ }\href@noop {} {\bibfield  {journal} {\bibinfo  {journal} {Opt. Express}\ }\textbf {\bibinfo {volume} {20}},\ \bibinfo {pages} {22819} (\bibinfo {year} {2012})}\BibitemShut {NoStop}%
\bibitem [{\citenamefont {Ye}\ \emph {et~al.}(2022)\citenamefont {Ye}, \citenamefont {Lei}, \citenamefont {Twayana}, \citenamefont {Girardi}, \citenamefont {Andrekson},\ and\ \citenamefont {Torres-Company}}]{ye2022integrated}%
  \BibitemOpen
  \bibfield  {author} {\bibinfo {author} {\bibfnamefont {Z.}~\bibnamefont {Ye}}, \bibinfo {author} {\bibfnamefont {F.}~\bibnamefont {Lei}}, \bibinfo {author} {\bibfnamefont {K.}~\bibnamefont {Twayana}}, \bibinfo {author} {\bibfnamefont {M.}~\bibnamefont {Girardi}}, \bibinfo {author} {\bibfnamefont {P.~A.}\ \bibnamefont {Andrekson}},\ and\ \bibinfo {author} {\bibfnamefont {V.}~\bibnamefont {Torres-Company}},\ }\bibfield  {title} {\bibinfo {title} {Integrated, ultra-compact high-q silicon nitride microresonators for low-repetition-rate soliton microcombs},\ }\href@noop {} {\bibfield  {journal} {\bibinfo  {journal} {Laser Photonics Rev.}\ }\textbf {\bibinfo {volume} {16}},\ \bibinfo {pages} {2100147} (\bibinfo {year} {2022})}\BibitemShut {NoStop}%
\bibitem [{\citenamefont {Liang}\ \emph {et~al.}(2021)\citenamefont {Liang}, \citenamefont {Bai}, \citenamefont {Yan}, \citenamefont {Wang}, \citenamefont {Zhang},\ and\ \citenamefont {Jin}}]{liang2021comprehensive}%
  \BibitemOpen
  \bibfield  {author} {\bibinfo {author} {\bibfnamefont {C.}~\bibnamefont {Liang}}, \bibinfo {author} {\bibfnamefont {Q.}~\bibnamefont {Bai}}, \bibinfo {author} {\bibfnamefont {M.}~\bibnamefont {Yan}}, \bibinfo {author} {\bibfnamefont {Y.}~\bibnamefont {Wang}}, \bibinfo {author} {\bibfnamefont {H.}~\bibnamefont {Zhang}},\ and\ \bibinfo {author} {\bibfnamefont {B.}~\bibnamefont {Jin}},\ }\bibfield  {title} {\bibinfo {title} {A comprehensive study of optical frequency domain reflectometry},\ }\href@noop {} {\bibfield  {journal} {\bibinfo  {journal} {IEEE Access}\ }\textbf {\bibinfo {volume} {9}},\ \bibinfo {pages} {41647} (\bibinfo {year} {2021})}\BibitemShut {NoStop}%
\bibitem [{\citenamefont {Finot}\ \emph {et~al.}(2008)\citenamefont {Finot}, \citenamefont {Kibler}, \citenamefont {Provost},\ and\ \citenamefont {Wabnitz}}]{finot2008beneficial}%
  \BibitemOpen
  \bibfield  {author} {\bibinfo {author} {\bibfnamefont {C.}~\bibnamefont {Finot}}, \bibinfo {author} {\bibfnamefont {B.}~\bibnamefont {Kibler}}, \bibinfo {author} {\bibfnamefont {L.}~\bibnamefont {Provost}},\ and\ \bibinfo {author} {\bibfnamefont {S.}~\bibnamefont {Wabnitz}},\ }\bibfield  {title} {\bibinfo {title} {Beneficial impact of wave-breaking for coherent continuum formation in normally dispersive nonlinear fibers},\ }\href@noop {} {\bibfield  {journal} {\bibinfo  {journal} {JOSA B}\ }\textbf {\bibinfo {volume} {25}},\ \bibinfo {pages} {1938} (\bibinfo {year} {2008})}\BibitemShut {NoStop}%
\bibitem [{\citenamefont {Dudley}\ and\ \citenamefont {Coen}(2002)}]{dudley2002coherence}%
  \BibitemOpen
  \bibfield  {author} {\bibinfo {author} {\bibfnamefont {J.~M.}\ \bibnamefont {Dudley}}\ and\ \bibinfo {author} {\bibfnamefont {S.}~\bibnamefont {Coen}},\ }\bibfield  {title} {\bibinfo {title} {Coherence properties of supercontinuum spectra generated in photonic crystal and tapered optical fibers},\ }\href@noop {} {\bibfield  {journal} {\bibinfo  {journal} {Opt. Lett.}\ }\textbf {\bibinfo {volume} {27}},\ \bibinfo {pages} {1180} (\bibinfo {year} {2002})}\BibitemShut {NoStop}%
\bibitem [{\citenamefont {Rebolledo-Salgado}\ \emph {et~al.}(2022)\citenamefont {Rebolledo-Salgado}, \citenamefont {Ye}, \citenamefont {Christensen}, \citenamefont {Lei}, \citenamefont {Twayana}, \citenamefont {Schr{\"o}der}, \citenamefont {Zelan},\ and\ \citenamefont {Torres-Company}}]{rebolledo2022coherent}%
  \BibitemOpen
  \bibfield  {author} {\bibinfo {author} {\bibfnamefont {I.}~\bibnamefont {Rebolledo-Salgado}}, \bibinfo {author} {\bibfnamefont {Z.}~\bibnamefont {Ye}}, \bibinfo {author} {\bibfnamefont {S.}~\bibnamefont {Christensen}}, \bibinfo {author} {\bibfnamefont {F.}~\bibnamefont {Lei}}, \bibinfo {author} {\bibfnamefont {K.}~\bibnamefont {Twayana}}, \bibinfo {author} {\bibfnamefont {J.}~\bibnamefont {Schr{\"o}der}}, \bibinfo {author} {\bibfnamefont {M.}~\bibnamefont {Zelan}},\ and\ \bibinfo {author} {\bibfnamefont {V.}~\bibnamefont {Torres-Company}},\ }\bibfield  {title} {\bibinfo {title} {Coherent supercontinuum generation in all-normal dispersion si3n4 waveguides},\ }\href@noop {} {\bibfield  {journal} {\bibinfo  {journal} {Opt. Express}\ }\textbf {\bibinfo {volume} {30}},\ \bibinfo {pages} {8641} (\bibinfo {year} {2022})}\BibitemShut {NoStop}%
\bibitem [{\citenamefont {Seok}\ \emph {et~al.}(2019)\citenamefont {Seok}, \citenamefont {Kwon}, \citenamefont {Henriksson}, \citenamefont {Luo},\ and\ \citenamefont {Wu}}]{seok2019wafer}%
  \BibitemOpen
  \bibfield  {author} {\bibinfo {author} {\bibfnamefont {T.~J.}\ \bibnamefont {Seok}}, \bibinfo {author} {\bibfnamefont {K.}~\bibnamefont {Kwon}}, \bibinfo {author} {\bibfnamefont {J.}~\bibnamefont {Henriksson}}, \bibinfo {author} {\bibfnamefont {J.}~\bibnamefont {Luo}},\ and\ \bibinfo {author} {\bibfnamefont {M.~C.}\ \bibnamefont {Wu}},\ }\bibfield  {title} {\bibinfo {title} {Wafer-scale silicon photonic switches beyond die size limit},\ }\href@noop {} {\bibfield  {journal} {\bibinfo  {journal} {Optica}\ }\textbf {\bibinfo {volume} {6}},\ \bibinfo {pages} {490} (\bibinfo {year} {2019})}\BibitemShut {NoStop}%
\bibitem [{\citenamefont {Zhang}\ \emph {et~al.}(2022)\citenamefont {Zhang}, \citenamefont {Kwon}, \citenamefont {Henriksson}, \citenamefont {Luo},\ and\ \citenamefont {Wu}}]{zhang2022large}%
  \BibitemOpen
  \bibfield  {author} {\bibinfo {author} {\bibfnamefont {X.}~\bibnamefont {Zhang}}, \bibinfo {author} {\bibfnamefont {K.}~\bibnamefont {Kwon}}, \bibinfo {author} {\bibfnamefont {J.}~\bibnamefont {Henriksson}}, \bibinfo {author} {\bibfnamefont {J.}~\bibnamefont {Luo}},\ and\ \bibinfo {author} {\bibfnamefont {M.~C.}\ \bibnamefont {Wu}},\ }\bibfield  {title} {\bibinfo {title} {A large-scale microelectromechanical-systems-based silicon photonics {L}i{DAR}},\ }\href@noop {} {\bibfield  {journal} {\bibinfo  {journal} {Nature}\ }\textbf {\bibinfo {volume} {603}},\ \bibinfo {pages} {253} (\bibinfo {year} {2022})}\BibitemShut {NoStop}%
\bibitem [{\citenamefont {Zhang}\ \emph {et~al.}(2021)\citenamefont {Zhang}, \citenamefont {Gu}, \citenamefont {Jiang}, \citenamefont {Thompson}, \citenamefont {Cai}, \citenamefont {Paesani}, \citenamefont {Santagati}, \citenamefont {Laing}, \citenamefont {Zhang}, \citenamefont {Yung} \emph {et~al.}}]{zhang2021optical}%
  \BibitemOpen
  \bibfield  {author} {\bibinfo {author} {\bibfnamefont {H.}~\bibnamefont {Zhang}}, \bibinfo {author} {\bibfnamefont {M.}~\bibnamefont {Gu}}, \bibinfo {author} {\bibfnamefont {X.}~\bibnamefont {Jiang}}, \bibinfo {author} {\bibfnamefont {J.}~\bibnamefont {Thompson}}, \bibinfo {author} {\bibfnamefont {H.}~\bibnamefont {Cai}}, \bibinfo {author} {\bibfnamefont {S.}~\bibnamefont {Paesani}}, \bibinfo {author} {\bibfnamefont {R.}~\bibnamefont {Santagati}}, \bibinfo {author} {\bibfnamefont {A.}~\bibnamefont {Laing}}, \bibinfo {author} {\bibfnamefont {Y.}~\bibnamefont {Zhang}}, \bibinfo {author} {\bibfnamefont {M.}~\bibnamefont {Yung}}, \emph {et~al.},\ }\bibfield  {title} {\bibinfo {title} {An optical neural chip for implementing complex-valued neural network},\ }\href@noop {} {\bibfield  {journal} {\bibinfo  {journal} {Nat. Commun.}\ }\textbf {\bibinfo {volume} {12}},\ \bibinfo {pages} {457} (\bibinfo {year} {2021})}\BibitemShut {NoStop}%
\bibitem [{\citenamefont {Gao}\ \emph {et~al.}(2025)\citenamefont {Gao}, \citenamefont {Sun}, \citenamefont {Rebolledo-Salgado}, \citenamefont {Van~Laer}, \citenamefont {Torres-company},\ and\ \citenamefont {Jochen}}]{gao_2025_14606023}%
  \BibitemOpen
  \bibfield  {author} {\bibinfo {author} {\bibfnamefont {Y.}~\bibnamefont {Gao}}, \bibinfo {author} {\bibfnamefont {Y.}~\bibnamefont {Sun}}, \bibinfo {author} {\bibfnamefont {I.}~\bibnamefont {Rebolledo-Salgado}}, \bibinfo {author} {\bibfnamefont {R.}~\bibnamefont {Van~Laer}}, \bibinfo {author} {\bibfnamefont {V.}~\bibnamefont {Torres-company}},\ and\ \bibinfo {author} {\bibfnamefont {S.}~\bibnamefont {Jochen}},\ }\href {https://doi.org/10.5281/zenodo.14606023} {\bibinfo {title} {Data accompanying 'tightly-confined and long z-cut lithium niobate waveguide with ultralow-loss'}} (\bibinfo {year} {2025})\BibitemShut {NoStop}%
\end{thebibliography}%

\end{document}